\documentclass{aastex}
\usepackage{emulateapj5,epsfig}

\def\be{\begin{equation}}
\def\ee{\end{equation}}
\def\ba{\begin{eqnarray}}
\def\ea{\end{eqnarray}}
\def\go{\mathrel{\raise.3ex\hbox{$>$}\mkern-14mu
             \lower0.6ex\hbox{$\sim$}}}
\def\lo{\mathrel{\raise.3ex\hbox{$<$}\mkern-14mu
             \lower0.6ex\hbox{$\sim$}}}
\def\cJ{{\cal J}}

\begin{document}

\title{Magnetically Driven Precession of Warped Disks and Milli-Hertz
Variabilities in Accreting X-Ray Pulsars}
\author{Akiko Shirakawa and Dong Lai}
\affil{Center for Radiophysics and Space Research, Cornell University,
Ithaca, NY 14853\\
Email: shirak,dong@astro.cornell.edu}

%%%%%%%%%%%%%%%%%%%%%%%%%%%%%%%%%%%%%%%%%%%%%%%%%%%%%%%%%%%%%%%%%
\begin{abstract}
The inner region of the accretion disk around a magnetized neutron star is
subjected to magnetic torques that induce warping and precession of the 
disk. These torques arise from interactions between the stellar field and
the induced electric currents in the disk. We carry out a global analysis 
of warping/precession modes in a viscous accretion disk, 
and show that under a wide range of conditions typical of accreting X-ray 
pulsars, the magnetic warping torque can overcome viscous damping and 
make the mode grow. The warping/precession modes are concentrated near the
inner edge of the disk (at the magnetosphere-disk boundary), and can give rise to
variabilities or quasi-periodic oscillations (QPOs) in the X-ray/UV/optical
fluxes from X-ray pulsars. We examine the observed properties  
of mHz QPOs in several systems (such as 4U 1626-67), and suggest that 
some hitherto unexplained QPOs are the results of magnetically 
driven disk warping/precession.
\end{abstract}

\keywords{accretion, accretion disks -- stars: neutron -- 
stars: magnetic fields -- pulsars -- binaries: close}

%%%%%%%%%%%%%%%%%%%%%%%%%%%%%%%%%%%%%%%%%%%%%%%%%%%%%%%%%%%%%%%%%
\section{Introduction}

Disk accretion onto a magnetic star occurs in a variety of astrophysical 
contexts, including accreting neutron stars, white dwarfs and 
pre-main-sequence stars (e.g., Frank et al.~1992). The basic
picture of disk--magnetosphere interaction is well-known: at large radii the
disk is unaffected by the stellar magnetic field; a somewhat sudden
transition occurs when the stellar field disrupts the disk at the
magnetospheric boundary (where the magnetic and plasma stresses balance), and
channels the plasma onto the polar caps of the star. 
A large number of theoretical papers have been written on the subject of 
the interaction between accretion disks and magnetized stars
(e.g., Pringle \& Rees 1972; Ghosh \& Lamb 1979,~1992; Aly 1980; 
Anzer \& B\"orner 1980,~1983; Lipunov et al.~1981; 
Wang 1987,~1995; Aly \& Kuijpers 1990; Spruit \& Taam 1993; Shu et al.~1994; 
van Ballegooijen 1994; Lovelace et al.~1995,1998; 
Li, Wickramasinge \& R\"udiger 1996; Campbell 1997; Lai 1998;
Terquem \& Papaloizou 2000), but numerical study of this problem
is still in its infancy (e.g., Hayashi et al.~1996;
Miller \& Stone 1997; Goodson et al.~1997; Fendt \& Elstner 2000).
Outstanding issues remain, including the efficiency of field
dissipation in/outside the disk, whether the disk excludes the stellar 
field by diamagnetic currents or the field can penetrate a large 
fraction of the disk, whether the threaded field remains closed (connecting the
star and the disk) or becomes open by differential shearing, and whether/how
magnetically driven wind is launched from the disk or the
magnetosphere/corotation boundary. 
 
Many previous theoretical papers have, for simplicity, adopted
the idealized geometry in which the magnetic axis, the spin axis and 
the disk angular momentum are aligned. However, in Lai (1999), it was 
shown that under quite general conditions, the stellar magnetic
field can induce warping in the inner disk and make the disk 
precess around the spin axis (see \S 2). Such magnetically driven warping 
and precession open up new possibilities for the dynamical behaviors of 
disk accretion onto magnetic stars. In Shirakawa \& Lai (2001) we have 
studied these effects in weakly magnetized accreting neutron stars and 
showed that the magnetic warping/precession effects may explain 
several observed features of low-frequency quasi-periodic oscillations 
in low-mass X-ray binaries.

In this paper we study global, magnetically driven warping/precession 
modes of inner disks of highly magnetized ($B\sim 10^{12}$~G) neutron stars
(NSs), as in accreting X-ray pulsars (\S 2 and \S 3). Such study is the first
step toward understanding the observational manifestations of the
magnetic warping/precession effects. We are 
motivated by the observations of the milli-Hertz quasi-periodic oscillations
(QPOs) in a number of X-ray pulsars (see Table 1 in \S 4 and references
therein). Of particular interest is the recent detection of 
the 1~mHz optical/UV oscillations in 4U 1626-67 (Chakrabarty et al.~2001). 
%are most like due to warping of the inner disk
In \S 4, we use the results of \S 3 and suggest that magnetically 
driven disk warping and precession naturally explain this and some other mHz
variabilities observed in X-ray pulsars.

%%%%%%%%%%%%%%%%%%%%%%%%%%%%%%%%%%%%%%%%%%%%%%%%%%%%%%%%%%%%%%%%%
\section{Magnetically Driven Warping/Precession and its Global Modes}

The inner region of the accretion disk onto a rotating magnetized central star
is subjected to magnetic torques that induce warping and precession of the 
disk (Lai 1999). These magnetic torques result from the interactions between
the accretion disk and the stellar magnetic field.
Depending on how the disk responds to the stellar field,
two different kinds of torque arise:
(i) If the vertical stellar magnetic field $B_z$ penetrates the disk,
it gets twisted by the disk rotation
to produce an azimuthal field $\Delta B_\phi=\mp\zeta B_z$ that has different
signs above and below the disk ($\zeta$ is the azimuthal pitch of the field
line and depends on the dissipation in the disk), and a radial surface current
$K_r$ results. The interaction between $K_r$ and the stellar $B_\phi$ gives
rise to a vertical force. While the mean force (averaging over the azimuthal
direction) is zero, the uneven distribution of the force induces a net
{\it warping torque} which tends to misalign the angular momentum of the disk
with the stellar spin axis.
(ii) If the disk does not allow the vertical stellar field
(e.g., the rapidly varying component of $B_z$ due to stellar rotation)
to penetrate, an azimuthal screening current $K_\phi$ will be induced on the
disk. This $K_\phi$ interacts with the radial magnetic field $B_r$
and produces a vertical force. The resulting {\it precessional torque}
tends to drive the disk into retrograde precession around the stellar spin
axis.

In general, both the magnetic warping torque and the precessional torque are present.
For small disk tilt angle $\beta$ (the angle between the disk normal and the
spin axis), the precession angular frequency and warping rate
at radius $r$ are given by
(see Lai 1999)\footnote{Note that the stellar spin frequency $\Omega_s$ 
does not appear in eqs.~(1) \& (2) since the variation of the field geometry 
due to the spin has been averaged out; this is justified because 
$\Omega_s\gg |\Omega_p|,~|\Gamma_w|$.}
\ba
&&\Omega_p (r)=\frac{\mu^2}{\pi^2 r^7\Omega(r)\Sigma(r) D(r)}F(\theta),
\label{eqn:Omega_p}\\
&&\Gamma_w (r)=\frac{\zeta\mu^2}{4\pi r^7\Omega(r)\Sigma(r)}\cos^2\theta,
\label{eqn:Gamma_w}
\ea
where $\mu$ is the stellar magnetic dipole moment, $\theta$ is
the angle between the magnetic dipole axis and the spin axis,
$\Omega(r)$ is the orbital angular frequency, and $\Sigma(r)$ is the surface
density of the disk. The dimensionless function $D(r)$ is given by
\be
D(r)={\rm max}~\left(\sqrt{r^2/r^2_{\rm in}-1}, \sqrt{2H(r)/r_{\rm in}}\right)
\label{eqn:D(r)},
\ee
where $H(r)$ is the half-thickness and $r_{\rm in}$ is the inner radius of the disk.
The function $F(\theta)$ depends on the dielectric property of the disk. We can write
\be
F(\theta)=2f\cos^2\theta-\sin^2\theta,
\ee
so that $F(\theta)=-\sin^2\theta$ if only the spin-variable vertical field is
screened out by the disk ($f=0$), and $F(\theta)=3\cos^2\theta-1$ if all vertical
field is screened out ($f=1$). In reality, $f$ lies between 0 and 1.
For concreteness, we shall set $F(\theta)=-\sin^2\theta$ in the following.

For accretion-powered X-ray pulsars, the inner disk radius $r_{\rm in}$ is
given by the magnetosphere radius $r_m$:
\be
r_{\rm in}\equiv\eta\left(\frac{\mu^4}{GM\dot M^2}\right)^{1/7}
=(3.4\times10^8~{\rm cm})\,\eta\,\mu_{30}^{4/7}M_{1.4}^{-1/7}{\dot M}_{17}^{-2/7},
\label{eqn:r_m}
\ee
where $M=(1.4M_{\odot})M_{1.4}$ is the neutron star mass,
${\dot M}=(10^{17}\,{\rm g\,s^{-1}}){\dot M_{17}}$ is
the mass accretion rate, $\mu_{30}=\mu/(10^{30}~{\rm G}\,{\rm cm}^3)$,
and $\eta\sim0.5-1$.
For typical parameters, the precession frequency is
\ba
\frac{\Omega_p(r)}{2\pi}&=&-(11.8~{\rm mHz})\,\mu_{30}^2M_{1.4}^{-1/2}
r_8^{-11/2}\nonumber\\
&&\times \left[\frac{\Sigma(r)}{10^4~{\rm g\,cm}^{-2}}\right]^{-1}
\left[\frac{D(r)}{0.1}\right]^{-1}\sin^2\theta, \label{eqn:Omega_p2}
\ea
where we have used $\Omega(r)=(GM/r^3)^{1/2}$, and $r_8=r/(10^8~{\rm cm})$.
The warping rate $\Gamma_w(r)$ is of the same order of magnitude as 
$\Omega_p(r)$.

Since the precession rate $\Omega_p(r)$ depends strongly on $r$,
coupling between different rings is needed to produce a global coherent
precession. The coupling can be achieved either by viscous stress or
through bending waves (e.g., Papaloizou\& Pringle~1983; Papaloizou \& Terquem~1995;
Larwood et al.~1996; Terquem~1998). In the viscosity dominated regime
(i.e., the dimensionless viscosity parameter $\alpha$ greater than $H/r$),
the dynamics of the warps can be studied using the formalism of
Papaloizou \& Pringle (1983) (see also Pringle~1992; Ogilvie~1999;
Ogilvie \& Dubus~2001). We model the disk as a collection of
rings which interact with each other via viscous stresses.
Each ring at radius $r$ has the unit normal vector ${\bf\hat l}(r,t)$.
In the Cartesian coordinates, with the $z$-axis along the neutron star spin,
we write
${\hat{\bf l}}=(\sin\beta\cos\gamma,\sin\beta\sin\gamma,\cos\beta)$,
with $\beta(r,t)$ the tilt angle and $\gamma(r,t)$ the twist angle.
For $\beta\ll 1$, the dynamical warp equation for ${\hat{\bf l}}$
(Lai 1999; see Papaloizou \& Pringle 1983; Pringle 1992) reduces to
an equation for $W(r,t)\equiv \beta(r,t)e^{i\gamma(r,t)}$:
\ba
&&\frac{\partial W}{\partial t}-
\left[\frac{3\nu_2}{4r}\left(1+\frac{2r\cJ'}{3\cJ}\right)
+\frac{3\nu_1}{2r}(\cJ^{-1}-1)\right]
\frac{\partial W}{\partial r}\nonumber\\
&&\qquad\qquad =\frac{1}{2}\nu_{2}\frac{\partial^2 W}{\partial r^2}
+i\Omega_pW+\Gamma_wW,\label{eqn:evolution}
\ea
where $\cJ'=d\cJ/dr$, $\nu_1$ is the usual viscosity 
, and $\nu_2$ is the viscosity which tends to reduce the disk tilt.
We assume that the ratio of $\nu_2$ to $\nu_1$ is constant.
In deriving eq.~(\ref{eqn:evolution}), we have used the relations for the
radial velocity and surface density: $v_r=-3\nu_1\cJ^{-1}/2r$ and
$\Sigma={\dot M}\cJ/3\pi\nu_1$.
The values and functional forms of $\nu_1$, $\nu_2$, $\Omega_p$,
$\Gamma_w$ [see eqs.~(\ref{eqn:Omega_p}) \& (\ref{eqn:Gamma_w})],
and the dimensionless function $\cJ(r)$ [see eq.~(\ref{eqn:J(r)}) below]
depend on disk models.

\subsection{Power-law Disk Models}
To gain insight on the properties of the global warping-precessional modes,
we consider power-law disk models, with $\Sigma(r)\propto r^\mu$.
We also assume that $D(r)$= constant, $\cJ(r)=1$,
$\nu_2/\nu_1=1$, and $\Omega(r)\propto r^{-3/2}$ as in the Keplerian flow.
Then, from eqs.~(\ref{eqn:Omega_p}) \& (\ref{eqn:Gamma_w}), we have
\ba
&&\Omega_p(r)=\Omega_p(r_{\rm in})
\left(\frac{r}{r_{\rm in}}\right)^{-\mu-11/2},\\
&&\Gamma_w(r)=\Gamma_w(r_{\rm in})
\left(\frac{r}{r_{\rm in}}\right)^{-\mu-11/2}.
\ea
Using the relation $\Sigma={\dot M}\cJ/3\pi\nu_1$,
the viscosity rate $\tau_{\rm visc}^{-1}$ can be written as
\be
\tau_{\rm visc}^{-1}(r)\equiv\frac{\nu_2(r)}{r^2}
=\tau_{\rm visc}^{-1}(r_{\rm in})\left(\frac{r}{r_{\rm in}}\right)^{-\mu-2}.
\ee
We look for a solution of the form
$W(r,t)=e^{i\sigma t}W(r)$ with the complex mode frequency $\sigma$ ($=\sigma_r+i\sigma_i$).
It is convenient to define dimensionless quantities by
\ba
&&\hat\sigma\equiv\sigma(r_{\rm in})\tau_{\rm visc}(r_{\rm in}),~~~
\hat\Omega_p\equiv\Omega_p(r_{\rm in})\tau_{\rm visc}(r_{\rm in}),\nonumber\\
&&\hat\Gamma_w\equiv\Gamma_w(r_{\rm in})\tau_{\rm visc}(r_{\rm in}).
\label{eqn:dimensionless}
\ea
Equation (\ref{eqn:evolution}) then reduces to the dimentionless form:
\be
i{\hat\sigma}W-\frac{3}{4x^{\mu+1}}\frac{dW}{dx}
=\frac{1}{2x^\mu}\frac{d^2W}{dx^2}
+i\frac{\hat\Omega_p}{x^{\mu+11/2}}W
+\frac{\hat\Gamma_w}{x^{\mu+11/2}}W,\label{eqn:oureqn1}
\ee
where $x\equiv r/r_{\rm in}$.

It is clear from eq.~(\ref{eqn:oureqn1}) that, for a given
$\mu$, the mode frequency $\hat\sigma$ depends only on
two dimensionless parameters $\hat\Omega_p$ and $\hat\Gamma_w$.
To solve eq.~(\ref{eqn:oureqn1}) for the complex eigenfunction $W(x)$ and
eigenvalue ${\hat \sigma}$, six real boundary conditions are needed.
In our calculation, the disk extends from $x_{\rm in}=1$ to $x_{\rm out}=50$.
For large $x$ and large $|\hat \sigma|$,
equation (\ref{eqn:oureqn1}) can be solved analytically, giving
\be
W(x)\propto\exp\,\left[\frac{2\sqrt{2}}{\mu+2}\,(i{\hat\sigma})^{1/2}
x^{\mu/2+1}\right],
\ee
where the sign of $(i{\hat \sigma})^{1/2}$ should be chosen so that
$W(x)\rightarrow 0$ as $x\rightarrow\infty$.
This approximate analytical solution, evaluated at $x_{\rm out}$,
together with its derivative, gives four (real) outer boundary conditions.
The inner boundary condition generally takes the form $W'(x_{\rm in})=aW(x_{\rm in})$,
with $a$ being a constant. Most of our results in \S 3 are based on $a=0$
(corresponding to zero torque at the inner edge of the disk), although we have
experimented with different $a$'s and found that for $|a|\lo1$
similar results are obtained (see Shirakawa \& Lai~2001).
In numerically searching a mode, we make a guess
for the eigenvalue ${\hat \sigma}$ and
integrate eq.~(\ref{eqn:oureqn1}) from $x_{\rm out}$ to $x_{\rm in}$.
Since $W(x)$ changes very rapidly from $x_{\rm out}$ to $x_{\rm in}$, we
rewrite eq.~(\ref{eqn:oureqn1}) in terms of a new function $w$ defined
as $W=e^w$ and use that equation for integration.
We find the correct value of ${\hat\sigma}$ that satisfies the boundary conditions
using the globally convergent Newton method (Press et al. 1992).

\subsection{Middle-Region Solution of the $\alpha$-Disk}
Here we consider the ``middle-region'' (gas-pressure -and scattering-dominated)
solution of the $\alpha$-disk
(Shakra \& Sunyaev 1973; Novikov \& Thorne 1973) which is relevant to
the inner part of the disk in accretion-powered X-ray pulsars.
In this model,
\ba
\Sigma&=&(7.5\times 10^3\,{\rm g}\,{\rm cm}^{-2})\alpha_{-1}^{-4/5}
M_{1.4}^{1/5}{\dot M}_{17}^{3/5}r_8^{-3/5}\!\cJ^{3/5},
\label{eqn:surface density}\\
H&=&(1.0\times 10^6\,{\rm cm})\alpha_{-1}^{-1/10}M_{1.4}^{-7/20}
{\dot M}_{17}^{1/5}r_8^{21/20}\cJ^{1/5},
\label{eqn:half-thickness}
\ea
where $\alpha=(0.1)\alpha_{-1}$ is the $\alpha$-viscosity parameter.
%and $r_8=r/(10^8\,{\rm cm})$.
The dimensionless function $\cJ(r)$ is given by
\footnote{Magnetic fields threading the disk can modify $\cJ(r)$ in a model-dependent
way (see Lai~1999 for an example). However, the basic feature can still be
approximated by eq. (\ref{eqn:J(r)}).}
\be
\cJ(r)=1-\xi\sqrt{\frac{r_{\rm in}}{r}}\label{eqn:J(r)},
\ee
where $\xi$ is a dimensionless parameter with
$0\leq\xi<1$ ($\xi=0$ corresponds to to zero net angular momentum transfer
across the inner disk, i.e., when the star is in spin equilibrium).
Substituting equation (\ref{eqn:surface density}) into
equations (\ref{eqn:Omega_p}) and (\ref{eqn:Gamma_w}), and using
$\Omega(r)=(GM/r^3)^{1/2}$, we get
\ba
\Omega_p (r)&=&(-9.8\times 10^{-3}~{\rm s}^{-1})\,\sin^2\theta\mu_{30}^2
\alpha_{-1}^{4/5}M_{1.4}^{-7/10}\nonumber\\
&&\times {\dot M}_{17}^{-3/5}
r_8^{-49/10}\cJ(r)^{-3/5}D(r)^{-1},\label{eqn:Omega_p2}\\
\Gamma_w(r)&=&(7.7\times 10^{-3}~{\rm s}^{-1})\,\zeta\cos^2\theta\mu_{30}^2
\alpha_{-1}^{4/5}M_{1.4}^{-7/10}\nonumber\\
&&\times {\dot M}_{17}^{-3/5} r_8^{-49/10}\cJ(r)^{-3/5}.
\label{eqn:Gamma_w2}
\ea
Using $\nu_1=\alpha H^2\Omega$, the viscosity rate is calculated as
\ba
\tau_{\rm visc}^{-1}(r)\equiv{\nu_2(r)\over r^2}
&=&(1.4\times10^{-4}~{\rm s}^{-1})\,\left(\frac{\nu_2}{\nu_1}\right)\,
\alpha_{-1}^{4/5}M_{1.4}^{-1/5}\nonumber\\
&&\times {\dot M_{17}}^{2/5}r_8^{-7/5}\cJ(r)^{2/5}.
\label{eqn:visc_rate}
\ea
With $W(r,t)=e^{i\sigma t}W(r)$, and using the dimensionless quantities
$\hat\sigma$, $\hat\Omega_p$, and $\hat\Gamma_w$
as defined in eq.~(\ref{eqn:dimensionless}),
we can write equation (\ref{eqn:evolution}) in the dimensionless form:
\ba
&&i{\hat\sigma}W-\left[\frac{3}{4}
\left(1+\frac{2x\cJ'}{3\cJ}\right)
+\frac{3\nu_1}{2\nu_2}\left(\frac{1}{\cJ}-1\right)\right]
\!\!\left(\frac{\cJ}{x\cJ_{\rm in}}\right)^{2/5}\!\!\frac{dW}{dx}\nonumber\\
&&=\frac{x^{3/5}}{2}\frac{\cJ^{2/5}}{\cJ_{\rm in}^{2/5}}
\frac{d^2W}{dx^2}+i\frac{\hat\Omega_p}{x^{4.9}}
\frac{D_{\rm in}\cJ_{\rm in}^{3/5}}{D\cJ^{3/5}}W
+\frac{\hat\Gamma_w}{x^{4.9}}\frac{\cJ_{\rm in}^{3/5}}{\cJ^{3/5}}W,\label{eqn:oureqn2}
\ea
where $D_{\rm in}\equiv D(r_{\rm in})$, $\cJ_{\rm in}\equiv \cJ(r_{\rm in})$, 
and $\cJ'=d\cJ/dx$. Using eqs.~(\ref{eqn:D(r)}), (\ref{eqn:r_m}),
and (\ref{eqn:half-thickness}), we can calculate $D_{\rm in}$ as
\be
D_{\rm in}=0.14\left(\frac{\eta}{0.5}\right)^{1/40}\!\!
\mu_{30}^{1/70}\alpha_{-1}^{-1/20}M_{1.4}^{-5/28}\dot M_{17}^{13/140}\cJ_{\rm in}^{1/10}
.\label{eqn:D_{in}}
\ee
Using eqs.~(\ref{eqn:r_m}) \& (\ref{eqn:D_{in}}) in eqs.~(\ref{eqn:Omega_p2}),
(\ref{eqn:Gamma_w2}), \& (\ref{eqn:visc_rate}), we can calculate
$\hat\Omega_p$ and $\hat\Gamma_w$ as
\ba
\hat\Omega_p
&=&-38.0\,\left(\frac{\nu_1}{\nu_2}\right)\left(\frac{\eta}{0.5}\right)^{-141/40}
\left(\frac{\sin^2\theta}{0.5}\right)\mu_{30}^{-1/70}\alpha_{-1}^{1/20}
\nonumber\\
&&\qquad \times
M_{1.4}^{5/28}{\dot M_{17}}^{-13/140}\cJ_{\rm in}^{-11/10},
\label{eqn:hatOmega_p}\\
\hat\Gamma_w
&=&21.4\,\left(\frac{\nu_1}{\nu_2}\right)\left(\frac{\eta}{0.5}\right)^{-7/2}
\left(\frac{\zeta}{5}\right)\left(\frac{\cos^2\theta}{0.5}\right)
\cJ_{\rm in}^{-1}.
\label{eqn:hatGamma_w}
\ea
Note that under the assumptions of $\cJ(x)=1$ and $D(x)=D_{\rm in}$,
eq.~(\ref{eqn:oureqn2}) reduces to eq.~(\ref{eqn:oureqn1}) with $\mu=-0.6$.

%%%%%%%%%%%%%%%%%%%%%%%%%%%%%%%%%%%%%%%%%%%%%%%%%%%%%%%%%%%%%%%%%
\section{Numerical Results}

\subsection{Mode Eigenfunction and Eigenvalue}
We first consider the power-law disk models of \S2.1.
For a given set of parameters ($\mu$, $\hat\Omega_p$, $\hat\Gamma_w$),
equation (\ref{eqn:oureqn1}) allows for many eigenmodes.
We shall focus on the ``fundamental'' mode which is more concentrated
near the inner edge of the disk and has larger $\hat\sigma_r$
(global precession frequency) and smaller $\hat\sigma_i$ (damping rate)
than any other ``higher-order'' modes.
%(see Fig.~1)

Figure 1 shows the tilt angle $\beta(x,t=0)=|W(x)|$ associated with the modes
for different sets of ($\hat\Omega_p$, $\hat\Gamma_w$), all with $\mu=-0.6$
(corresponding to the ``middle-region'' $\alpha$-disk with $\cJ(x)=1$
and $D(x)=D_{\rm in}$).
We see that as $|\hat\Omega_p|$
(note $\hat\Omega_p<0$ due to retrograde precession)
and $\hat\Gamma_w$ increase,
the modes become more concentrated near the inner radius of the disk.
This behavior can be understood heuristically: for a given $|\Omega_p(r_{\rm in})|$,
a larger $|\hat\Omega_p|$ implies smaller viscosity [see eq.~(\ref{eqn:dimensionless})],
and thus the coupling between different disk radii is reduced, and the mode is
less spread.

Figure~2(a) shows the magnitude of the mode frequency ($\hat\sigma_r<0$ due
to retrograde precession)
$|\hat\sigma_r|$ in units of $|\hat\Omega_p|$
as a function of $|\hat\Omega_p|$ for different values of $\hat\Gamma_w$.
The ranges of $|\hat\Omega_p|$ and $\hat\Gamma_w$ are chosen to be
from 10 to 1000 to cover possible values of parameters for
accretion-powered X-ray pulsars [see eqs.~(\ref{eqn:hatOmega_p})
\& (\ref{eqn:hatGamma_w}) with $\cJ_{\rm in}=1$].
We include the $\mu=-1.0$ and $\mu=1.0$ results as well as the
``middle region'' $\mu=-0.6$ result
to show how our results vary with the change of the surface density power-law.
We see that $|\hat\sigma_r/\hat\Omega_p|=|\sigma_r/\Omega_p(r_{\rm in})|$
always lies between 0.3 to 0.85.
The ratio $|\hat\sigma_r/\hat\Omega_p|$ increases as
$|\hat\Omega_p|$ and $\hat\Gamma_w$ increase. This is consistent with the behavior
of the mode eigenfunction (see Fig.~1) that a larger $|\hat\Omega_p|$
and $\hat\Gamma_w$ make the mode more concentrated near the inner disk edge.

\subsection{Global Warping Instability Criterion}
In the absence of magnetic warping torque ($\Gamma_w=0$), we expect
the disk warp to be damped by the viscous stress acting on the differential
precession ($\Omega_p$). This ``magnetic Bardeen-Petterson effect'' (Lai~1999)
is analogous to the usual Bardeen-Petterson effect (Bardeen \& Petterson~1975),
where the combined effects of viscosity and differential Lense-Thirring
precession align the rotation axis of the inner disk with the spin axis of the
rotating black hole (or rotating, non-magnetic NS).
The competition between the magnetically driven warping ($\Gamma_w$)
and the magnetic Bardeen-Petterson damping can be determined by our
global analysis.

Figure~2(b) shows the damping rate $\hat\sigma_i$ as a function of $|\hat\Omega_p|$
for different values of $\hat\Gamma_w$.
We see that $\hat\sigma_i$ decreases as $\hat\Gamma_w$ increases, and
becomes negative (implying mode growth) when the ratio
$\hat\Gamma_w/|\hat\Omega_p|$ is sufficiently large.
By solving eq.~(\ref{eqn:oureqn1}) with $\mu=-0.6$, we find that the numerical
value of $\hat\sigma_i$ can be approximated by
\be
\hat\sigma_i=-a\hat\Gamma_w+b|{\hat\Omega_p}|^{0.6},
\ee
with $a\sim (0.5-1.0)$ and $b\sim (0.5-1.0)$; this result is insensitive
to modest change of the surface density power-law ($\mu=-1$ to 1).
For the mode to grow ($\hat\sigma_i<0$) we require
\be
\hat\Gamma_w\go2|\hat\Omega_p|^{0.6}~\Longleftrightarrow~{\rm Global~Warping~Instability}.
\label{eqn:criteria}
\ee
Using eqs.~(\ref{eqn:hatOmega_p}) and (\ref{eqn:hatGamma_w}), this condition becomes
\ba
&& 1.6\left(\frac{\nu_1}{\nu_2}\right)^{0.4}\!\!
\left(\frac{\eta}{0.5}\right)^{-1.39}\!\!
\left(\frac{\zeta}{5}\right)\cos^2\theta\,(\sin^2\theta)^{-0.6}
\mu_{30}^{0.0086}\nonumber\\
&&\qquad \times \alpha_{-1}^{-0.03}
M_{1.4}^{-0.11}\dot M_{17}^{0.056}\cJ_{\rm in}^{-0.34} \go 1
\label{eqn:criteria2}
\ea
We see that for parameters which characterize X-ray pulsars
the mode growth condition can be satisfied, although not always.
In general, high (but not unreasonable) $\zeta$ ($>$ a few) and
small $\cJ_{\rm in}$ (see \S 3.3) are preferred to obtain growing modes.

\subsection{Effect of $\cJ(r)$}
The results of \S3.1--3.2 are based on eq.~(\ref{eqn:oureqn1}), corresponding
to eq.~(\ref{eqn:oureqn2}) with $D(x)=D_{\rm in}$ and $\cJ(x)=1$.
Here we consider the solutions of
eq.~(\ref{eqn:oureqn2}) with the function $\cJ(x)$ given by eq.~(\ref{eqn:J(r)}).

Figure~3 shows the mode frequency $\hat\sigma$ as a function of
$\cJ_{\rm in}=1-\xi$ [see eq.~(\ref{eqn:J(r)})] for three
different sets of ($\hat\Omega_p$, $\hat\Gamma_w$):
($-38\cJ_{\rm in}^{-1.1}$, $21\cJ_{\rm in}^{-1}$),
($-7.6\cJ_{\rm in}^{-1.1}$, $38\cJ_{\rm in}^{-1}$), and
($-38\cJ_{\rm in}^{-1.1}$, $43\cJ_{\rm in}^{-1}$); these are obtained
from eqs.~(\ref{eqn:hatOmega_p}) \& (\ref{eqn:hatGamma_w}) with
($\sin^2\theta$, $\zeta$)=(0.5, 5), (0.1, 5), and (0.5, 10), respectively,
while other parameters being fixed to the standard values [$\eta=0.5$, $\nu_2/\nu_1=1$,
$\mu_{30}=1$, $\alpha_{-1}=1$, $M_{1.4}=1$, $\dot M_{17}=1$].
In the calculations, we set $D(x)=D_{\rm in}$ for simplicity;
using the more accurate function $D(x)$ given in eq.~({\ref{eqn:D(r)}}) 
only slightly changes the numerical results.
We see from Fig.~3 that $|\hat\sigma_r/\hat\Omega_p|$ is insensitive
to the choice of $\xi$ since the most of the dependence on $\cJ_{\rm in}$
is already absorbed into the definition of $\hat\Omega_p$ [see eq.~(\ref{eqn:hatOmega_p})].
We also see that a small $\cJ_{\rm in}$ tends to increase $\hat\sigma_i/|\hat\Omega_p|$,
although growing warping modes still exist for a wide range of
parameters.
We find that the simple global warping instability criteria
given in eqs.~(\ref{eqn:criteria}) and (\ref{eqn:criteria2})
can be used for $\cJ_{\rm in}>0.1$.
For smaller $\cJ_{\rm in}$ ($\lo 0.1$), $\hat\sigma_i$ should be
obtained numerically to determine whether the mode grows or gets damped.

%%%%%%%%%%%%%%%%%%%%%%%%%%%%%%%%%%%%%%%%%%%%%%%%%%%%%%%%%%%%%%%%
\section{Applications to Milli-Hertz QPO's in Accreting X-ray Pulsars}

QPOs with frequencies $1-100$~mHz have been detected in at least 11 
accreting X-ray pulsars (Table~1; see also Boroson et al.~2000). 
%These include high-mass giant/supergiant systems, transient Be-binary systems,
%and low-mass X-ray binaries. 
%Although the nature of the mHz QPOs is still unclear, it is generally 
%believed to be related to the interactions between the magnetosphere 
%of the neutron star and the inner part of the accretion disk.
These mHz QPOs are often interpreted in terms of the 
beat frequency model (BFM; Alpar \& Shaham~1985; Lamb et al.~1985)
or the Keplerian frequency model (KFM; van der Klis et al.~1987). 
In the BFM, the observed QPO frequency represents the beat 
between the Keplerian frequency $\nu_K$ at the inner disk radius 
$r_{\rm in}$ and the NS spin frequency $\nu_s$ [i.e., $\nu_{\rm QPO}
=\nu_K(r_{\rm in})-\nu_s$]. In the KFM, the QPOs arise from the 
modulation of the X-rays by some inhomogeneities in the inner disk 
at the Keplerian frequency [i.e., $\nu_{\rm QPO}=\nu_K(r_{\rm in})$].
However, we see from Table~1 that for several sources, more than
one QPOs have been detected and the difference in the QPO frequencies
is not equal to the spin frequency. Thus it is evident that 
the KFM and/or the BFM
cannot be the whole story.  We also note that in both the KFM and the BFM, 
it is always postulated that the inner disk contain some blobs or 
inhomogeneities, whose physical origin is unclear.

%Table~1 lists the physical parameters of the 11 X-ray binary pulsars for
%which QPOs with frequencies $1-100$~mHz have been detected.

Here we suggest a ``Magnetic Disk Precession Model'' (MDPM) for the 
mHz variabilities and QPOs of accreting X-ray pulsars.
The magnetically driven precession of the warped inner disk (outside
but close to the magnetosphere boundary) can modulate
X-ray/UV/optical flux in several ways: (i) The observed radiation
(in UV and optical, depending on $r_{\rm in}$) from the inner
disk due to intrinsic viscous dissipation varies as the 
angle between the local disk normal vector and the line-of-sight changes
during the precession; (ii) The flux of reprocessed UV/optical disk
emission visible along our sight also varies as the reprocessing geometry 
evolves; (iii) Modulation of the X-ray flux arises from regular 
occulation/obscuration of the radiation from the central NS and magnetosphere
by the precessing inner disk. In the MDPM, we identify $\nu_{\rm QPO}$ 
with the global precession frequency driven by the magnetic
torques. Our calculations in \S3 show that under a wide range of conditions,
the warping/precession mode is concentrated near the disk inner edge, and
the global mode frequency is equal to $A=0.3-0.85$ (depending on details of 
the disk structure) times the magnetically driven precession frequency at
$r_{\rm in}=r_m$. Thus we write $\nu_{\rm QPO}=A|\Omega_p(r_{\rm in})|/2\pi$.
Using eq.~(\ref{eqn:Omega_p2}) together with eqs.~(\ref{eqn:r_m}),
(\ref{eqn:surface density}), and (\ref{eqn:D_{in}}), we have
(for the $\alpha$-disk model),
\ba
\nu_{\rm QPO}
&=&(15.7~{\rm mHz})A\,\sin^2\theta\,\mu_{30}^2\alpha_{-1}^{4/5}M_{1.4}^{-7/10}
\dot M_{17}^{-3/5}\nonumber\\
&&\qquad \times\left(\frac{r_{\rm in}}{10^8~{\rm cm}}\right)^{-49/10}\!
\left(\frac{D_{\rm in}}{0.1}\right)^{-1}\!\cJ_{\rm in}^{-3/5}
\nonumber\\
&=&(0.83~{\rm mHz})\,A\,
\left(\frac{\eta}{0.5}\right)^{-4.93}\!\!
\sin^2\theta\,\mu_{30}^{-0.81}\nonumber\\
&&\qquad \times \alpha_{-1}^{0.85}M_{1.4}^{0.18}\dot M_{17}^{0.71}
\cJ_{\rm in}^{-0.7}.
\label{eqn:compare}
\ea
We also note that the Keplerian frequency at 
$r=r_{\rm in}=r_m$ is
%%% CHECK M^{5/7} not M^{3/14}
\be
\nu_K(r_{\rm in})=(985~{\rm mHz})\left(\frac{\eta}{0.5}\right)^{-3/2}\!\!
\mu_{30}^{-6/7}M_{1.4}^{5/7}\dot M_{17}^{3/7}.
\label{eqn:Kepler}
\ee
We see from eq.~(\ref{eqn:compare}) that the MDPM can produce QPOs
with frequencies of order $1$~mHz; larger values of
$\nu_{\rm QPO}$ would require $\cJ_{\rm in}\ll 1$ (corresponding to low surface
density at $r_{\rm in}$). The value of $\cJ_{\rm in}$ depends on 
details of the physics at the inner edge of the disk, therefore is 
uncertain. Let $V_r(r_{\rm in})=\chi c_s(r_{\rm in})$, where $V_r=\dot M/(2\pi
r\Sigma)$ is the radial velocity and $c_s=H\Omega$ is the sound speed of the
disk, we find
\be
\cJ_{\rm in}\simeq 3.1\times 10^{-4}\chi^{-5/4}\alpha_{-1}^{9/8}
M_{1.4}^{-7/16}\dot M_{17}^{1/4}\left({r_{\rm in}\over 10^8~{\rm
cm}}\right)^{1/16}.
\label{eq:cj}\ee
Setting $\chi=1$ would give the minimum value of $\cJ_{\rm in}$.

We now discuss several individual sources.
% for which estimates of $\mu_{30}$ (based on
% cyclotron line features) and $\dot M_{17}$ are available (see Table~1).

\subsection{4U 1626+67}

The LMXB 4U 1626+67 consists of a $\nu_s=130$~mHz 
X-ray pulsar in a 42 min orbit with very low-mass ($\lo 0.1\,M_\odot$)
companion (see Chakrabarty 1998). QPOs at $48$~mHz (and oscillations at
130~mHz) have been detected simultaneously in X-rays 
and in the optical/UV band (Shinoda et al.~1990; Chakrabarty~1998; 
Chakrabarty et al.~2001). Thus it is natural
to attribute the optical/UV variability to the reprocessing of the variable
X-ray emission in the accretion disk. Since $\nu_s>48$~mHz, the 
KFM is problematic because the propeller effect would inhibit 
accretion when $\nu_s>\nu_K(r_m)$. (It is still possible to ascribe the
QPO to Keplerian motion at some radius farther out in the disk, 
but this is rather ad hoc.) The BFM is a viable alternative for the 48~mHz
QPO. 

Recent HST observations by Chakrabarty et al.~(2001) revealed 
a strong QPO around 1~mHz (centroid frequency in the range of 0.3--1.2~mHz,
and $Q=\nu/\Delta\nu$ of order 10) in the optical/UV band. This QPO is absent
in simultaneous X-ray data, and is stronger in UV and weaker in the optical 
band. These features can be naturally explained as due to 
warping of the inner accretion disk (see Chakrabarty et al.~2001).
Indeed, using $B\simeq 3\times 10^{12}$~G (from BeppoSAX observations of a 
cyclotron feature; Orlandini et al.~1998) and $\dot M_{17}\simeq 1$
(corresponding to X-ray luminosity $L_X\simeq 10^{37}$~erg/s;
Chakrabarty 1998), eq.~(\ref{eqn:compare}) yields
$\nu_{\rm QPO}\simeq 0.34\,A\,(\eta/0.5)^{-4.9}
\sin^2\theta\,\alpha_{-1}^{0.85}\cJ_{\rm in}^{-0.7}$~mHz, 
which can easily give the observed $\nu_{\rm
QPO}=1$~mHz provided we allow for $\cJ_{\rm in}<1$ 
[see eq.~(\ref{eq:cj})]. If we interpret the 48~mHz QPO with the BFM, 
we have $\nu_K(r_{\rm in})=178$~mHz and $\eta\simeq 0.83$, we therefore
require $\cJ_{\rm in}\lo 0.01$ (depending on $\alpha$ and $\sin^2\theta$).

\subsection{Other Sources}

For the other sources listed in Table 1, no ``smoking-gun'' 
signature of warped disk has been observed. However, the numerical values of
QPO frequencies indicate that the MDPM may be at work.

{\it Her X--1: } This well-studied binary X-ray pulsar ($\nu_s=808$~mHz)
shows QPOs in the UV and optical bands at frequencies of $8\pm 2$ and 
$43\pm 2$~mHz; these QPOs most likely arise from the reprocessing
of the disk oscillations by the companion star (Boroson et al. 2000).
A QPO at 12~mHz in X-rays is also present, but its connection with the
8~mHz QPO is not clear (Moon \& Eikenberry 2001b). 
Since $\nu_s>\nu_{\rm QPO}$, the KFM is not applicable. The BFM predicts
$\nu_K(r_{\rm in})=816$~mHZ and 851~mHZ for $\nu_{\rm QPO}=8$~mHz and
43~mHz respectively. For $\mu_{30}\simeq 3$ (see Table~1)
, $\dot M_{17}\simeq 1$ (corresponding to $L_X\simeq 10^{37}~{\rm ergs}$; 
Choi et al.~1994), and $M_{1.4}\simeq 1$, we have $\nu_K(r_{\rm in})\simeq 
400\,(\eta/0.5)^{-3/2}$~mHz. Thus to explain the 8~mHz or 43~mHz QPO
with the BFM requires $\eta<0.5$. The phenomenology of the 12~mHz X-ray
QPO is consistent with disk precession (see Moon \& Eikenberry 2001b);
this may be explained by the MDPM.

{\it LMC X--4:} This persistent X-ray pulsar ($\nu_s=74$~mHz) exhibits large
X-ray flares. QPOs at frequencies of 0.65-1.35~mHz
and 2-20~mHz have been found during such flares (Moon \& Eikenberry 2001a).
Since $\nu_{s}>\nu_{\rm QPO}$, the KFM is not applicable.
The BFM predicts $\nu_K(r_m)\simeq 75$~mHz and $\simeq 76-94$~mHz
for $\nu_{\rm QPO}=0.65-1.35$~mHz and $2-20$~mHz, respectively.
Equation~(\ref{eqn:Kepler}) with the measured $\mu_{30}\simeq 11$ (see Table~1)
and $\dot M_{17}\simeq 25$ (corresponding to $L_X\simeq 5\times 10^{38}~{\rm ergs}$; 
Moon \& Eikenberry~2001c) gives
$\nu_K(r_{\rm in})=502\,(\eta/0.5)^{-3/2}$~mHz, which is 
much larger than the values of $\nu_K(r_{\rm in})$ required by the BFM
(even for $\eta=1$). Hence, it is difficult, if not impossible, 
to identify the observed QPO frequencies
(especially for $\nu_{\rm QPO}= 0.65-1.35$~mHz) with the beat frequency.
On the other hand, eq.~(\ref{eqn:compare}) gives
$\nu_{\rm QPO}\simeq 1.2\,A\,(\eta/0.5)^{-4.9}
\sin^2\theta\,\alpha_{-1}^{0.85}\cJ_{\rm in}^{-0.7}$~mHz,
which is close to the observed $\nu_{\rm QPO}=0.65-1.35$~mHz.

{\it 4U 0115+63:} This transient source shows a broad 62~mHz QPO during a
flaring state (Soong \& Swank 1989). Recent XTE observation also reveals
a prominent QPO in the X-ray flux at $2$~mHz (Heindl et al.~1999).
Heindl et al.~(1999) noted that this 2~mHz QPO may be explained by 
occultation of the radiation beam by the accretion disk. Analogous 
to the 1~mHz QPO of 4U 1626+67 (see \S 4.1), we suggest that magnetically
driven disk warping/precession may be responsible for such occultation.
With the well-measured magnetic field ($B\simeq 1.3\times 10^{12}$~G) 
of the NS from the cyclotron lines, similar constraints on the system
parameters can be obtained.

For other sources listed in Table 1, either only single QPO is known,
or there is no independent measurement/constraint on $\mu_{30}$ and
$\dot M$, thus the theoretical interpretation is currently ambiguous.

%%%%%%%%%%%%%%%%%%%%%%%%%%%%%%%%%%%%%%%%%%%%%%%%%%%%%%%%%%%%%%%%%
\section{Conclusions}

We have shown in this paper that the inner region of the disk around a 
strongly magnetized ($\sim 10^{12}$~G) neutron star can be warped and 
will precess around the stellar spin (see Lai 1999). These effects arise from 
the interactions between the stellar field and the induced currents in 
the disk (before it is disrupted at the magnetosphere boundary). 
We have carried out a global analysis of the warping/precession modes
and found that growing modes exist for a wide range of parameters 
typical of accreting X-ray pulsars. We therefore expect that the
magnetically driven warping/precession effect will give rise to 
variabilities/QPOs in the X-ray/UV/optical fluxes. 
We have suggested that some mHz QPOs observed in several systems (e.g., 4U
1626-67) are the results of these new magnetic effects. 
Although there are significant uncertainties in the physical conditions
of the magnetosphere-disk boundary [and these uncertainties
prohibit accurate calculation of the QPO frequency; see 
eq.~(\ref{eqn:compare})], we emphasize that the existence of these effects 
is robust. Continued study of the variabilities of X-ray pulsar systems
would provide useful constraints on the magnetosphere--disk interactions.

\acknowledgments
We thank D. Chakrabarty and W. Heindl for drawing our
attention to QPOs in 4U 1626-67 and 4U 0115+63.
This work is supported in part by NSF Grant AST 9986740,
as well as by a research fellowship (to D.L.)
from the Alfred P. Sloan foundation.

%%%%%%%%%%%%%%%%%%%%%%%%%%%%%%%%%%%%%%%%%%%%%%%%%%%%%%%%%%%%%%%%%%%%

%%%%%%%%%%%%%%%%%%%%%%%%%%%%%%%%%%%%%%%%%%%%%%%%%%%%%%%%%%%%%%%%%%%%
\bigskip
\bigskip
\bigskip
\clearpage
\begin{table}[h]
\caption{Accretion-powered X-ray pulsars with ${\rm mHz}$ QPOs}
\begin{tabular}{lcccc}
\hline
\hline
System & spin frequency & QPO frequency
& magnetic moment & References \tablenotemark{a}
\\
       & $\nu_s$~[mHz]  & $\nu_{\rm QPO}$~[mHz] & $\mu_{30}$\tablenotemark{b}
& \\

\hline
%{\it low-mass binaries}\\
4U 1626--67 & 130& 1, 48     & 3.3 & [1], [2], [3], $[4]^{\dagger}$\\
Her X--1    & 807.9& 8, 12, 43 & 3 & [5], [6], $[7]^{\dagger}$\\
%{\it high-mass supergiant and giant systems} &  & \\
SMC X-1 & 1410    & (60)\tablenotemark{c}&   & [8]\\
Cen X-3 & 207     & 35                    &    & [9]\\
LMC X-4 & 74      & 0.65--1.35, 2--20     & 11 & [10], $[11]^{\dagger}$\\
4U 1907+09 & 2.27 & 55                    & 2.5& [12], $[13]^{\dagger}$\\
EXO 2030+375 & 24  & 187--213 &    & [14]\\
4U 0115+63   & 277 & 2, 62    & 1.5& [15], [16]\\
XTE J1858+034& 4.5 & 111      &    & [17]\\
V 0332+53    & 229 & 51       & 2.5& [18], $[19]^{\dagger}$\\
A 0535+262   & 9.71& 27--72   & 9.5& [20], $[21]^{\dagger}$\\
\hline
\tablenotetext{a}{References: 
[1] Shinoda et al.~1990;
[2] Chakrabarty~1998;
[3] Chakrabarty et al.~2001;    
$[4]^{\dagger}$ Orlandini et al.~1998;
[5] Boroson et al.~2000; 
[6] Moon \& Eikenberry~2001;
$[7]^{\dagger}$; Makishima et al.~1999; 
[8] Wojdowski et al.~1998; 
[9] Takeshima et al.~1991; 
[10] Moon \& Eikenberry~2001;
$[11]^{\dagger}$ La Barbera et al.~2001; 
[12] in't Zand et al.~1998; 
$[13]^{\dagger}$ Makishima \& Mihara~1992;
[14] Angelini et al.~1989;
[15] Soong \& Swank~1989; 
[16] Heindl et al.~1999; 
[17] Paul \& Rao~1998; 
[18] Takeshima et al.~1994;
$[19]^{\dagger}$ Makishima et al.~1990; 
[20] Finger et al.~1996; 
$[21]^{\dagger}$ Grove et al.~1995. 
These references include QPO discovery (no mark) and 
magnetic field strength $B$ estimated from cyclotron features (with $\dagger$).
%and X-ray luminosity $L_X$ estimated from the X-ray flux and the source distance
%(with $\ddagger$).
}
\tablenotetext{b}{The magnetic moment is calculated by
$\mu=BR^3$, or $\mu_{30}=B_{12}R_6^3$. We assume $R=10$~km, or $R_6=1$ here.}
%\tablenotetext{c}{The mass accretion rate is calculated by
%$L_X=GM\dot M/R$, or $\dot M_{17}=5.3\,L_{X,\,38}M_{1.4}^{-1}R_6^{-1}$.
%Note that $L_X$ has uncertainties originated in the source distance
%and X-ray beaming factor.}
\tablenotetext{c}{marginal detection}
\end{tabular}
\end{table}
%%%%%%%%%%%%%%%%%%%%%%%%%%%%%%%%%%%%%%%%%%%%%%%%%%%%%%%%%%%%%%%%%%%%
\clearpage
\begin{figure}
\plotone{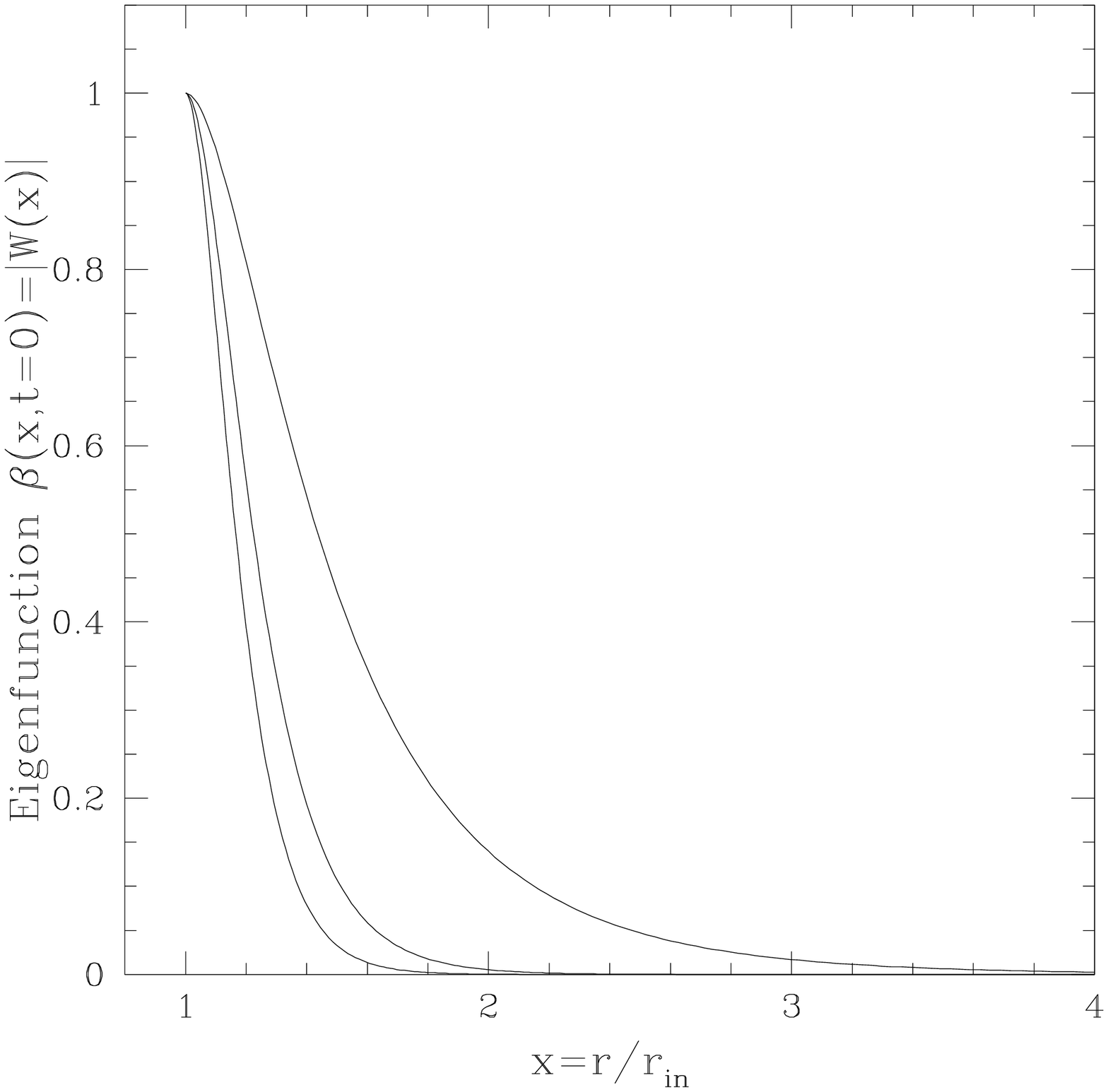}
\caption{The disk tilt angle of the warping/precession modes
as determined from eq.~(\ref{eqn:oureqn1}) with $\mu=-0.6$.
The curves represent the fundamental modes for
($\hat\Omega_p$, $\hat \Gamma_m$)
=$(-10, 100)$, $(-100, 10)$, and $(-10, 10)$ from left to right,
with the corresponding mode frequency
$\hat\sigma=(\hat\sigma_r,\hat\sigma_i)=(-6.9, -57)$, $(-62, 12)$, and $(-4.2, -1.7)$.
%The dotted curve represents a higher order mode
%for ($\hat \Omega_{LT}$, $\hat\Gamma_m$)=(100, 100),
%with $\hat\sigma=(\hat\sigma_r, \hat\sigma_i)=(41, 6.0)$.
The eigenfunction is normalized such that the maximum tilt angle is 1.
}
\end{figure}

\clearpage
\begin{figure}
\plotone{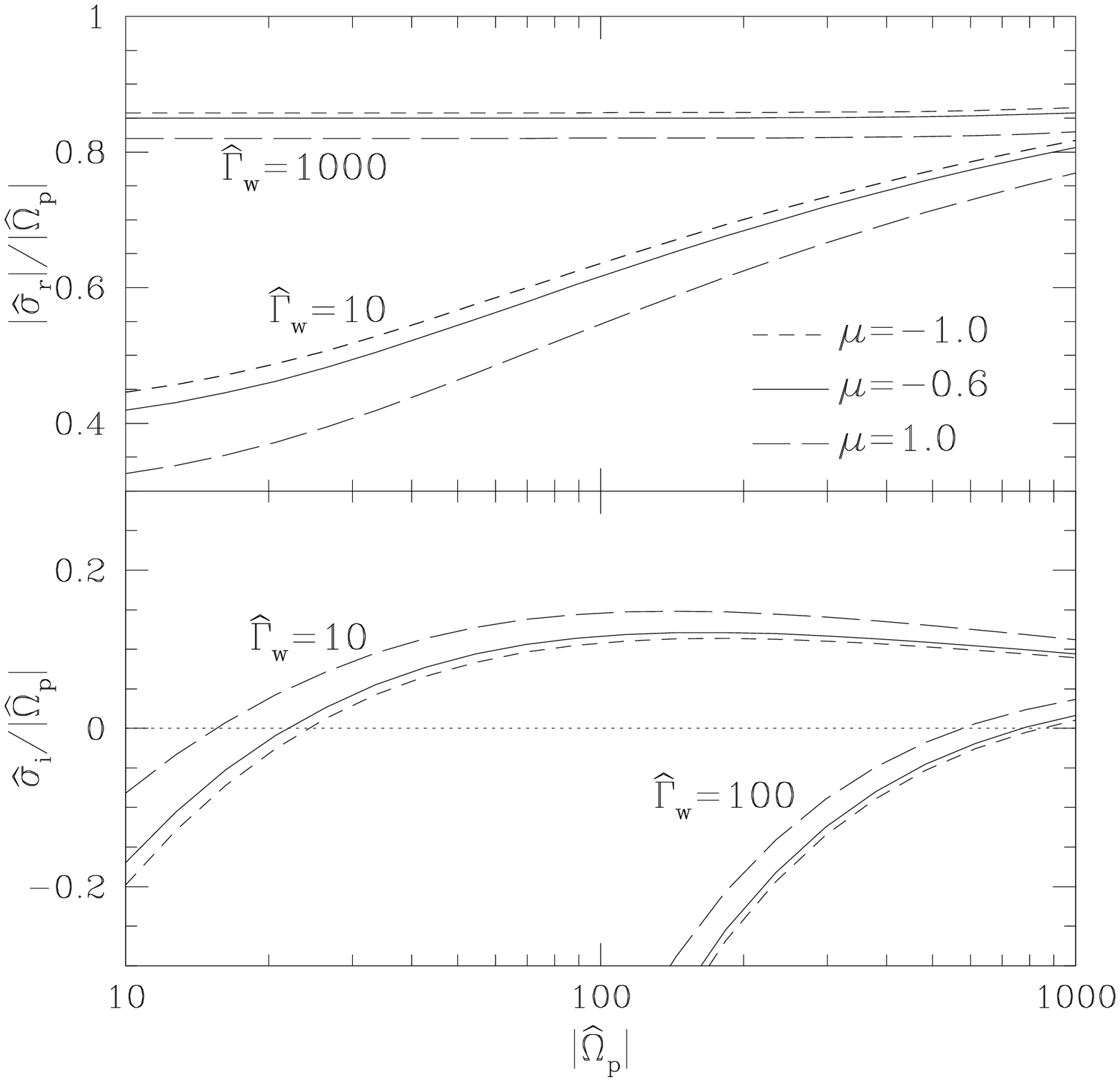}
\caption{
The upper panel shows the magnitude of the mode frequency $|\hat\sigma_r|$ in units of
$|\hat\Omega_p|$ as a function of $|\hat\Omega_p|$ for different values of $\hat
\Gamma_w$. Different surface density power-laws, $\Sigma\propto r^{\mu}$ with
$\mu=-1$, $-0.6$ and 1, are adopted [see eq.~(\ref{eqn:oureqn1})].
The lower panel shows the corresponding mode damping rate
$\hat\sigma_i$ (in units of $|\hat \Omega_p|$).
Note that negative $\hat\sigma_i$ implies growing mode.
}
\end{figure}

\clearpage
\begin{figure}
\plotone{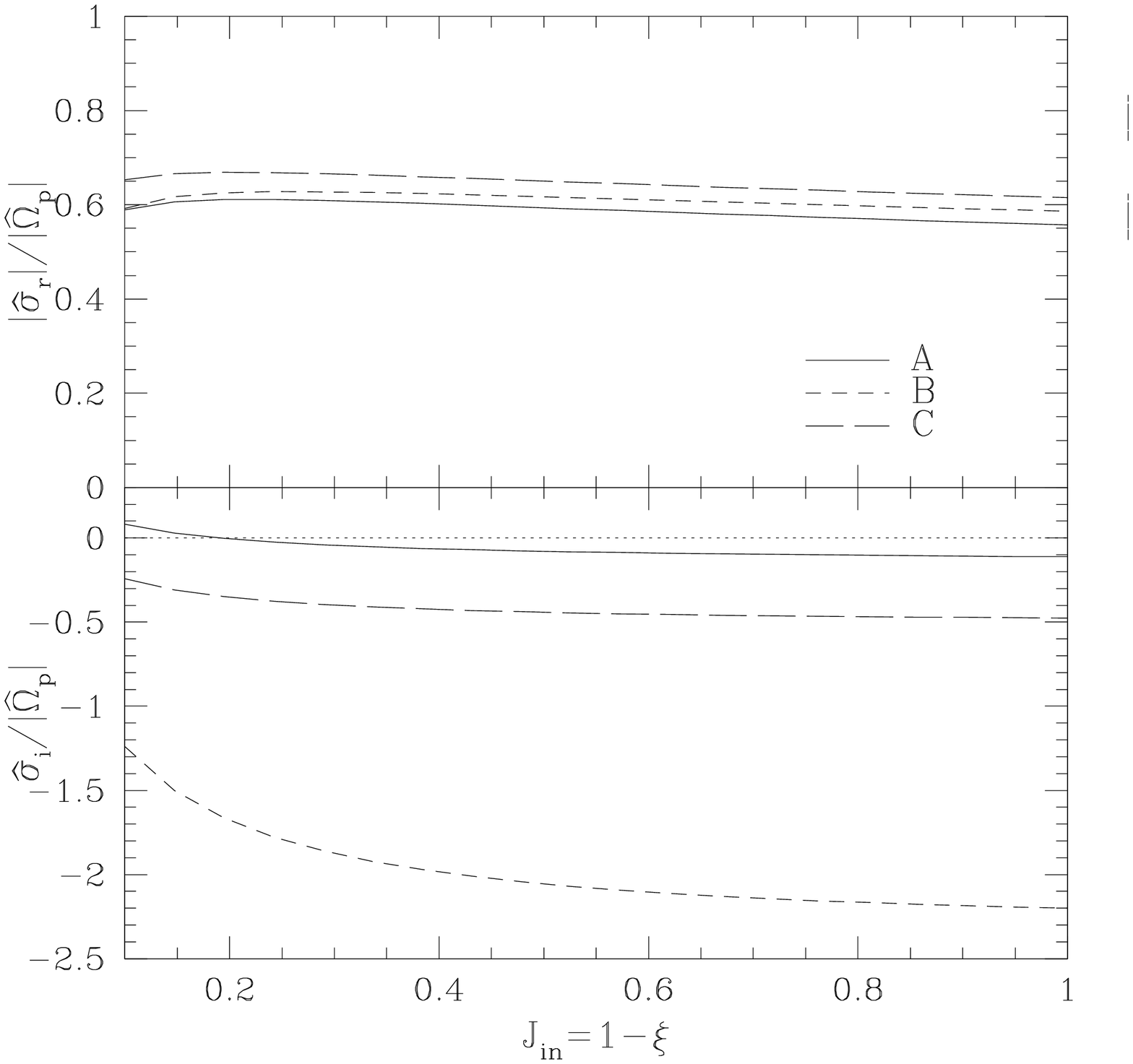}
\caption{
The upper panel shows the magnitude of the mode frequency $|\hat\sigma_r|$ in units of
$|\hat\Omega_p|$ as a function of $\cJ_{\rm in}=1-\xi$
for differet parameter sets A, B, and C: ($\hat\Omega_p$, $\hat\Gamma_w$)=
($-38\cJ_{\rm in}^{-1.1}$, $21\cJ_{\rm in}^{-1}$),
($-7.6\cJ_{\rm in}^{-1.1}$, $38\cJ_{\rm in}^{-1}$), and
($-38\cJ_{\rm in}^{-1.1}$, $43\cJ_{\rm in}^{-1}$), respectively (see the text).
The lower panel shows the corresponding mode damping rate
$\hat\sigma_i$ (in units of $|\hat \Omega_p|$).
Note that negative $\hat\sigma_i$ implies growing mode.
}
\end{figure}

\begin{thebibliography}{}

\bibitem[]{}
Alpar, M.~A., \& Shaham, J. 1985, Nature, 316, 239
%BFM paper

\bibitem[]{} 
Aly, J.~J. 1980, A\&A, 86, 192
% Exact solution of diamagnetic disk

\bibitem[]{} 
Aly, J.~J., \& Kuijpers, J. 1990, A\&A, 227, 473

\bibitem[]{}
Angelini, L., Stella, L., \& Parmar, A.~N. 1989, ApJ, 346, 906
%discovery of QPOs (drifting frequency?) in EXO 2030+375

\bibitem[]{} 
Anzer, U., \& B\"orner, G. 1980, A\&A, 83, 133

\bibitem[]{} 
Anzer, U., \& B\"orner, G. 1983, A\&A, 122, 73

%\bibitem[]{} 
%Armitage, P.~J., \& Clarke, C.~J. 1996, MNRAS, 280, 458
%Magnetic Breaking of T Tauri Stars

%bibitem[]{} 
%rons, J. 1987, in ``The Origin and Evolution of Neutron Stars'' (IAU
%ymp.~No.~125), ed. D.~J. Helfand \& J.-H. Huang (D. Reidel Pub.:
%ordrecht)

\bibitem[]{}
Bardeen, J.~M., \& Petterson, J.~A. 1975, ApJ, 195, L65

\bibitem[]{}
Boroson, B., et al. 2000, ApJ, 545, 399
%discovery of UV QPO in Her X--1

\bibitem[]{} 
Campbell, C.~G. 1997, Magnetohydrodynamics in Binary Stars
(Kluwer Academic Pub.: Dordrecht)

\bibitem[]{}
Chakrabarty, D. 1998, ApJ, 492, 342
%$L_X\simeq 10^{37}$~erg/s

\bibitem[]{}
Chakrabarty, D., et al. 2001, ApJ, in press (astro-ph/0106432)

\bibitem[]{}
Choi, C.~S., Nagase, F., Makino, F., Dotani, T., \& Min, K.~W. 1994, 
ApJ, 422, 799
%$L_X\simeq2.1\times 10^{37}$~erg/s for Her X--1

\bibitem[]{}
Fendt, C., \& Elstner, D. 2000, A\&A, 363, 208
% Long-term evolution of a dipole type magnetosphere interacting with an
% accretion disk. II. Transition into a quasi-stationary spherically radial
% outflow

\bibitem[]{}
Finger, M.~H., Wilson, R.~B., \& Harmon., B.~A. 1996, ApJ, 459, 288
%discovery of QPO in A0535+262

\bibitem[]{} 
Frank, J., King, A., \& Raine, D. 1992, Accretion Power in Astrophysics
(Cambridge Univ. Press)

\bibitem[]{} 
Ghosh, P., \& Lamb, F.~K. 1979, ApJ, 234, 296

\bibitem[]{}
Ghosh, P., \& Lamb, F.~K. 1992, in X-Ray Binaries and Recycled Pulsars,
ed. E.~P.~J. van den Heuval \& S.~A. Rappaport (Dordrecht: Kluwer), p.487

\bibitem[]{} 
Goodson, A.~P., Winglee, R.~M., \& B\"ohm, K.-H. 1997, ApJ, 489, 199
% Time-dependent accretion by magnetic young stellar objects as a launching
% mechanism for stellar jets

\bibitem[]{}
Grove J.~E., et al. 1995, ApJ, 438, L25
%$B\simeq 9.5\times10^{12}$~G for A0535+262

\bibitem[]{} 
Hayashi, M.~R., Shibata, K., \& Matsumoto, R. 1996, ApJ, 468, L37

\bibitem[]{} 
Heindl, W.~A., Coburn, W., Gruber, D.~E., Pelling, M.~R., \& Rothschild, R.~E. 1999, ApJ, 521, L49

%\bibitem[]{}
%in't Zand, J.~J.~M., Strohmayer, T.~E., Baykal, A. 1998, ApJ, 479, L47
%$L_X\simeq4\times 10^{36}$~erg/s during flares in 4U 1907+09

\bibitem[]{}
in't Zand, J.~J.~M., Baykal, A., \& Strohmayer, T.~E. 1998, ApJ, 496, 386
%discovery of QPOs in 4U 1907+09

%\bibitem[]{}
%Kendziorra, E., et al. 1994, A\&A, 291, L31
%$B\simeq 4.3\times10^{12}$~G for A0535+262 
%see also Grove et al.

%\bibitem[]{}
%Kommers, J.~M., Chakrabarty, D., \& Lewin, W.~H.~G.~1998, ApJ, 497, L33
%%sidebands of 4U 1626-67

\bibitem[]{}
La Barbera, A., Burderi, L., Di Salvo, T., Iaria, R., Robba, N.~R.
2001, ApJ, in press (astro-ph/0104367)
%$B\simeq 1.1\times10^{13}$~G for LMC X--4

%\bibitem[]{}
%Lai, D., Lovelace, R., Wasserman, I. 1999, unpublished, astro-ph/9904111

\bibitem[]{}
Lai, D. 1998, ApJ, 502, 721

\bibitem[]{}
Lai, D. 1999, ApJ, 524, 1030

\bibitem[]{}
Lamb, F.~K., Shibazaki, N., Alpar, M.~A., \& Shaham, J. 1985,
Nature, 317, 681
%BFM

\bibitem[]{}
Larwood, J.~D., Nelson, R.~P., Papaloizou, J.~C.~B., Terquem, C. 1996,
MNRAS, 282, L597

\bibitem[]{} 
Li, J., Wickramasinge, D.~T., \& R\"udiger, G. 1996, ApJ, 469, 765.
%Magnetized Accretion and Funnel Flow

\bibitem[]{} 
Lipunov, V.~M., Sem\"enov, E.~S., \& Shakura, N.~I. 1981, Sov. Astron., 25,
439.

\bibitem[]{} 
Lovelace, R.~V.~E., Romanova, M.~M., \& Bisnovatyi-Kogan, G.~S. 1995,
MNRAS, 275, 244.

\bibitem[]{} 
Lovelace, R.~V.~E., Romanova, M.~M., \& Bisnovatyi-Kogan, G.~S. 1999,
ApJ, 514, 368
% Magnetic Propeller Outflows

\bibitem[]{}
Makishima, K1990, ApJ, 365, L59
%$B\simeq 2.5\times10^{12}$~G for V 0332+53

\bibitem[]{}
Makishima, K., \& Mihara., T. 1992, in Proc. Yamada Conf.~28,
Frontiers of X-Ray Astronomy, ed. Y. Tanaka \& K. Koyama
(Tokyo: Universal Academy), 23
%$B\simeq 2.5\times10^{12}$~G for 4U 1907+09

\bibitem[]{}
Makishima, K., Mihara, T., Nagase, F., \& Tanaka, Y. 1999, ApJ, 525, 978
%$B\simeq 3\times10^{12}$~G for Her X--1

%\bibitem[]{}
%Markovi\'c, D., \& Lamb, F.~K. 1998, ApJ, 507, 316

\bibitem[]{} 
Miller, K.~A., \& Stone, J.~M. 1997, ApJ, 489, 890

\bibitem[]{}
Moon, D-.S., \& Eikenberry, S.~S. 2001a, ApJ, 549, L225
%discovery of QPOs in LMC X--4

\bibitem[]{}
Moon, D-.S., \& Eikenberry, S.~S. 2001b, ApJ, 552, L135
%discovery of X-ray QPO in Her X--1

\bibitem[]{}
Moon, D-.S., \& Eikenberry, S.~S. 2001c, ApJ, submitted
%$L_X\simeq 5\times 10^{38}$~erg/s for LMC X--4

\bibitem[]{}
Novikov, I.~D., \& Thorne, K.~S. 1973, in Black Holes, ed. C. DeWitt
\& B. DeWitt (New York: Gordon \& Breach), 343
%disk

\bibitem[]{}
Ogilvie, G.~I.~1999, MNRAS, 304, 557

\bibitem[]{}
Ogilvie, G.~I., \& Dubus, G. 2001, MNRAS, 320, 485

%\bibitem[]{}
%Orlandini, M., et al. 1997, in The Active
%X-ray Sky: Results from BeppoSAX and Rossi-XTE (astro-ph/9712023)
%$B=3.3\times 10^{12}$~G for 4U 1626-67

\bibitem[]{}
Orlandini, M., et al. 1998, ApJ, 500, 163
% 4U 1627 's B field from cyclotron line

\bibitem[]{}
Paul, B., \& Rao, A.~R.~1998, A\&A, 337, 815
%discovery of QPO in XTE J1858+034

\bibitem[]{}
Papaloizou, J.~C., \& Pringle, J.~E. 1983, MNRAS, 202, 1181
%disk eqn

\bibitem[]{}
Papaloizou, J.~C., \& Terquem, C. 1995, MNRAS, 274, 987

\bibitem[]{}
Press, W.~H., Teukolsky, S., Vetterling, W.~T., \& Flannery, B.~P.
1992, Numerical Recipes (Cambridge: Cambridge Univ. Press)

\bibitem[]{}
Pringle, J.~E. 1992, MNRAS, 258, 811
%disk eqn

%\bibitem[]{}
%Pringle, J.~E. 1996, MNRAS, 281, 857

\bibitem[]{} 
Pringle, J.~E., \& Rees, M.~J. 1972, A\&A, 21, 1

%\bibitem[]{}
%Psaltis, D., et al. 1999, ApJ, 520, 763

%\bibitem[]{}
%Psaltis, D., Belloni, T., \& van der Klis, M. 1999,
%ApJ, 520, 262

\bibitem[]{}
Shakura, N.~I., \& Sunyaev, R.~A. 1973, A\&A, 24, 337

\bibitem[]{}
Shinoda, K., Kii, T., Mitsuda, K., Nagase, F., Tanaka, Y., Makishima, K.,
\& Shibazaki, N. 1990, PASJ, 42, L27

\bibitem[]{}
Shirakawa, A., \& Lai, D.~2001, ApJ, in press

\bibitem[]{} 
Shu, F.~H., et al.~1994, ApJ, 429, 781

\bibitem[]{}
Soong, Y., \& Swank, J.~H. 1989, in Proc. 23rd ESLAB Symposium on Two
Topics in X-ray Astronomy, ed. N.~E. White, J.~J. Hunt, \& B. Battrick
(Paris: ESA), 617
%discovery of QPO in 4U 0115+63

\bibitem[]{} 
Spruit, H.~C., \& Taam, R.~E. 1993, ApJ, 402, 593

\bibitem[]{}
Takeshima, T., Dotani, T., Mitsuda, K., \& Nagai, F. 1991, PASJ, 43, L43
%discovery of QPOs in Cen X--3

\bibitem[]{}
Takeshima, T., Dotani, T., Mitsuda, K., \& Nagase, F. 1994, ApJ, 436, 871
%discovery of QPO in V 0322+53
%also $L_X\simeq3.1\times 10^{37}$~erg/s for it

\bibitem[]{}
Terquem, C.~E.~J.~M.~L.~J. 1998, ApJ, 509, 819

\bibitem[]{}
Terquem, C., \& Papaloizou, J.C.B. 2000, A\&A, 360, 1031
% The response of an accretion disc to an inclined dipole with application
% to AA Tau

\bibitem[]{} 
van Ballegooijen, A.~A. 1994, Space Science Rev., 68, 299

%\bibitem[]{}
%van der Klis, M., et al. 1985, Nature, 316, 225

\bibitem[]{}
van der Klis, M., et al. 1987, ApJ, 316, 411
%KFM paper

%\bibitem[]{}
%van der Klis, M. 1995, in X-Ray Binaries, ed. W.~H.~G. Lewin,
%5J. van Paradijs, \& E.~P.~J. van den Heuvel (Cambridge: Cambridge Univ. Press), 252

%\bibitem[]{}
%van der Klis, M. 2000, ARAA, 38, 717

\bibitem[]{} 
Wang, Y.-M. 1987, A\&A, 183, 257

\bibitem[]{} 
Wang, Y.-M. 1995, ApJ, 449, L153

%\bibitem[]{}
%Wijnands, R.~A.~D. et al. 1996, ApJ, 469, L5

%\bibitem[]{}
%Wijnands, R., \& van der Klis, M. 1998 Nature, 394, 344

\bibitem[]{}
Wojdowski, P., Clark, G.~W., Levine, A.~M.,
Woo, J.~W., \& Zhang, S.~N. 1998, ApJ, 502, 253
%discovery of (60)-mHz QPO in SMC X--1

%\bibitem[]{} 
%Yi, I. 1995, ApJ, 442, 768.
%% Magnetized accretion and the spin evolution of classical T Tauri stars

\end{thebibliography}
\end{document}